\documentstyle[aps,twocolumn,graphicx]{revtex}

\begin{document}
\title{Depleted Kondo Lattices.}
\author{F.F. Assaad }
\address{Institut f\"ur Theoretische Physik III, \\
   Universit\"at Stuttgart, Pfaffenwaldring 57, D-70550 Stuttgart, Germany. \\
   Max Plank institute for solid state research, Heisenbergstr. 1,
   D-70569, Stuttgart.}
\maketitle

\begin{abstract}
We consider a two dimensional Kondo lattice model with exchange $J$ and hopping 
$t$ in which three out of four impurity spins are removed in a regular way. 
At the particle-hole symmetric point the model may be studied with
auxiliary field quantum Monte Carlo methods without sign problems. 
To achieve the relevant energy scales on finite clusters, 
we introduce a simple method to reduce size effects by up to an order of
magnitude in temperature.  
In this model, a metallic phase survives up to arbitrarily low temperatures 
before being disrupted by magnetic fluctuations which open a gap in the 
charge sector. We study the formation of the 
heavy-electron state with emphasis  on a crossover scale $T^{\star}$ 
defined  by the maximum in the resistivity versus temperature curve.   
The behavior of
thermodynamic properties such as specific heat as well as spin and charge uniform 
susceptibilities are  studied  as  the temperature  varies  in a wide range 
across $T^{\star}$.
Within our accuracy $T{^\star}$ compares well to the Kondo scale of the related 
single impurity problem. 
Finally our QMC resuls are compared with mean-field approximations.  
\end{abstract}

\section{Introduction}
Kondo insulators as well as  heavy fermion materials \cite{Lee86,Aeppli92}
are believed to be described
by Kondo lattice models with antiferromagnetic exchange $J$ and hopping $t$.
In this framework, the Kondo insulator 
- exemplified by Ce$_3$Bi$_4$Pt$_3$ - corresponds to a 
special band filling  in which there is exactly one conduction electron per 
impurity spin. The origin of the charge and spin gaps is easy to understand
in the limit of large  $J/t$ where each conduction electron is trapped  by 
an impurity spin to form a Kondo singlet. 
Starting from this point there are two ways of generating  metallic states 
which will described the heavy electron  state. 
On one hand, one can remove conduction electrons thus leaving uncompensated 
impurity spins. In the limit $J/t  \rightarrow \infty $ the problem maps onto 
the $U \rightarrow \infty $ Hubbard model where the Kondo singlets are 
represented by empty sites and the bachelor - or unscreened - spins by  
electrons \cite{Lacroix85}. Thus, a metallic state is expected. 
On the other hand one can keep the number of conduction electrons constant 
and deplete the lattice of impurity spins. Starting from 
Ce$_3$Bi$_4$Pt$_3$ this amounts to replacing Ce by La \cite{Hundley90}.
Each missing impurity spin liberates a conduction electron. 
As can be seen easily in the strong coupling limit this conduction electron is
localized since its motion involves the breaking of adjacent Kondo
singlets. Hence a bound state with magnetic properties appears 
within the gap \cite{Schlottmann96}. 
If the depletion is {\it large} enough the induced localized levels may 
overlap coherently to form a metallic state.  Alternatively this metallic 
state where there are more conduction electrons than impurity spins may be 
viewed as evolving from dilute impurity limit as realized for example in 
Ce$_x$La$_1-x$Cu$_6$ \cite{Sumiyama86}. Note however that in the small (large)
$x$ limit the Ce (La) atoms are randomly substituted for  La (Ce)
in the  LaCu$_6$ (CeCu$_6$) crystal structure.

In the next section, we first consider within a mean-field
approximation \cite{Zhang00b} both
above describe routes to obtain metallic states. The paramagnetic solution 
of this  mean-field approximation is nothing but the  large-N (N corresponds
to the SU(N) symmetry of the impurity spins) saddle point 
\cite{Newns87,Hewson,Georges00}.
Here, we  concentrate on a special choice of depletion scheme for the localized 
spins. Three out of four impurity spins are removed in a regular way thus 
ensuring that in the limit $J/t \rightarrow \infty$ the ground state is metallic.
At  the mean-field level both routes to obtain a metallic state 
(doping and above describe depletion scheme)  lead
to the same metallic ground state: a Fermi-liquid with arbitrarily 
large effective mass (or arbitrarily low coherence temperature) as the 
coupling $J/t$ is  reduced. The mean-field 
solution has several pitfalls. For instance,  the imaginary part of the 
self-energy vanishes at all temperatures, the Kondo scale corresponds to 
a phase transition rather than to a crossover scale and charge fluctuations 
on impurity sites are allowed since constraints are imposed only on average. 

For depleted Kondo lattices it is possible to carry out 
auxiliary field quantum Monte
Carlo (QMC)  simulations without encountering the infamous sign problem.  
This comes with the restriction of particle-hole symmetry for 
the conduction band.  Nevertheless this gives us the opportunity to compare
{\it exact} results with mean-field approximations. 
There are two challenges to circumvent before
achieving  good results. The first  one is the  implementation of the 
exchange interaction $J$ along with the constraint of singly occupied 
impurity sites. This problem has been solved in \cite{Assaad99a} and extensively 
discussed in \cite{Capponi00}. The second challenge lies in controlling finite 
size effects  stemming  from the conduction electrons.  This is extremely
important in the framework of Kondo physics since screening  takes place
on an energy scale set by the Kondo temperature $T_K $ which is exponentially
small in the weak coupling limit. If the level spacing of the 
conduction  electrons exceeds this small energy scale, 
finite size effect will  dominate the result.
In section \ref{SizeB.chap} we present a simple method to reduce 
by up to an order magnitude 
in temperature the onset of size effects in thermodynamic
quantities (specific heat, spin and charge  susceptibilities)  for a
given lattice size. 

Section \ref{QMC_res} summarizes our numerical results. 
Ground state, thermodynamic
and transport properties  are computed   on lattice sizes
up to 500 sites.  We show that our simulations 
reproduce the typical form of the resistivity versus temperature curve
observed in heavy fermion materials: an initial rise as a function of 
decreasing  temperature 
originating from the onset of screening of the magnetic impurities
then followed by a sharp downturn signaling the onset of
coherence between individual screening clouds. Energy scales as a function
of coupling strength are computed.
The last section is devoted to conclusions and discussion of our results.  

\section{Models and mean-field approximation}
Our starting point is the two-dimensional  Kondo lattice model (KLM): 
\begin{equation}
\label{KLM_Ham}
  H( \{\vec{R} \} ) = -t \sum_{\langle \vec{i},\vec{j}  \rangle, \sigma } 
 \left(  c^{\dagger}_{\vec{i}, \sigma}   
         c_{\vec{j}, \sigma} + {\rm H.c.} \right) + 
    \sum_{\vec{R}} \vec{S}^{c}_{\vec{R}} \cdot 
    \vec{S}^{f}_{\vec{R}}.
\end{equation}
Here $\vec{i}$ runs over the lattice sites of a two dimensional 
square lattice: $ \vec{i}  = n \vec{a}_x + m \vec{a}_y $, $ n,m: 1 \cdots L$
with $\vec{a}_x,\vec{a}_y$ the lattice vectors.  The set of  lattice
points $ \{\vec{R} \} $ denote the positions of impurity spins. 
$c^{\dagger}_{\vec{i},\sigma} $ creates a conduction electron on site 
$ \vec{i} $ with z-component of spin $\sigma$ and 
$ \vec{S}^{f}_{\vec{R}} =\frac{1}{2} \sum_{\sigma,\sigma'}
f^{\dagger}_{\vec{R},\sigma}  \vec{\sigma}_{\sigma,\sigma'}
f_{\vec{R},\sigma'} $  are the spin 1/2 operators  with  $ \vec{\sigma}$
the Pauli spin matrices. A similar definition holds for 
$ \vec{S}^{c}_{\vec{R}} $. Since the KLM stems from the strong coupling
limit of the periodic  Anderson model \cite{Schrieffer66} 
the charge degrees of freedom
on the f-sites are frozen. Hence, a constraint of one electron  per
$f$-site holds for KLM.  

\begin{figure}
\includegraphics[width=0.45\textwidth]{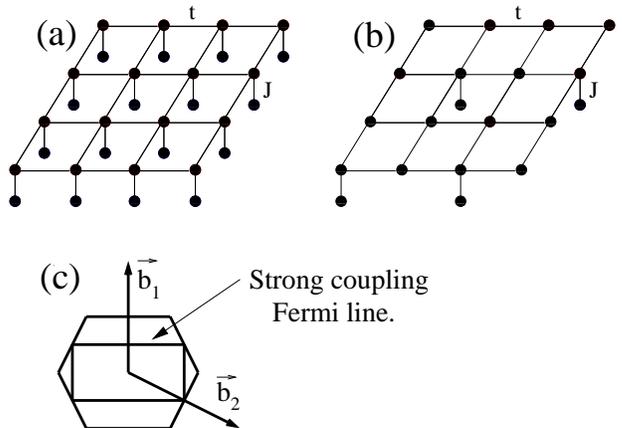}
\caption[]{(a) The standard KLM. The square lattice is spanned by the 
lattice vectors $\vec{a}_x$, $\vec{a}_y$. (b) The depleted KLM (DKLM).
The corresponding Bravais lattice is spanned by $2 \vec{a}_x $ and 
$\vec{a}_x + 2 \vec{a}_y $. Each unit cell contains a single localized orbital 
for the impurity spin and four delocalized orbitals accommodating the conduction
electrons.  
In both (a) and (b) the top layer denotes the conduction electrons with
hopping matrix element $t$. The bottom layer are the magnetic impurities 
which  couple to the conduction electrons via the exchange $J$.
(c) the Brillouin zone of the DKLM. The reciprocal lattice vectors read
$ \vec{b}_1 = \frac{\pi}{a}(0,1)$ and $ \vec{b}_2 = \frac{\pi}{a}(1,-1/2)$
}
\label{latt.fig}
\end{figure}

In the notation of Eq. \ref{KLM_Ham} the usual KLM model  is obtained
when $\vec{R} = n \vec{a}_x  +  m \vec{a}_y $ (see Fig. \ref{latt.fig}a). 
In principle, the depletion scheme of impurity spins can be random - a
situation which would reflect the experimental realization - or regular. As 
a first step we have chosen a regular depletion and considered
$ \vec{R} = n  \left[ 2 \vec{a}_x \right]  + 
m \left[  \vec{a}_x + 2 \vec{a}_y \right] $ (see Fig. \ref{latt.fig}b). 
From now onwards we will refer to this model as the depleted Kondo lattice
model (DKLM). The unit cell of the DKLM has five orbitals: one localized for
the impurity spin and four delocalized for the conduction electrons. The 
Bravais lattice is spanned by the lattice vectors: 
$\vec{a}_1 = 2 \vec{a}_x $ and
 $\vec{a}_2 = \vec{a}_x + 2 \vec{a}_y $.

Having defined the model, we 
now address the question: to which degree  does the KLM at 
$n_c < 1$ have the same physics as the DKLM at $n_c = 4$? Here 
$n_c$ corresponds to the number of conduction electrons per
unit cell.  As mentioned previously, 
in the strong coupling limit the KLM with band filling $n_c$
maps onto the $U \rightarrow 
\infty $ Hubbard model with band filling $\tilde{n}_c = 1 - n_c$
\cite{Lacroix85}. 
The nature of the metallic state  is not clear due to the strong 
correlations. However for fillings far from the doping induced Mott
metal-insulator transition, $\tilde{n}_c = 1$, it is reasonable to 
believe that beyond the one-dimensional case the ground state is a 
Fermi liquid as described for example
by a Gutzwiller approximation. 
Due to the special choice of the depletion, the strong coupling
DKLM at $n_c = 4$ 
is a Fermi liquid. In this limit, each impurity spin will capture a 
conduction electron and the remnant conduction electrons will form the
metallic state. To be more precise at $J/t \rightarrow \infty $
we obtain a three band model, where a single conduction band crosses 
the Fermi level, $\epsilon_F = 0$. The thus obtained strong coupling 
Fermi line is shown in Fig. \ref{latt.fig}c.

In the weak coupling limit, we can address the above question within a 
mean-field approximation \cite{Zhang00b}. 
One can write the KLM and DKLM models as:
\begin{eqnarray}
    & &  H( \{ \vec{R} \} ) =  \sum_{\vec{k},\sigma} \epsilon(\vec{k})
     c^{\dagger}_{\vec{k},\sigma} c_{\vec{k},\sigma} 
 \\
 & &  + \frac{J}{4} \sum_{\vec{R}}
         \left( f^{\dagger}_{\vec{R},\uparrow} f_{\vec{R},\uparrow} -
                f^{\dagger}_{\vec{R},\downarrow} f_{\vec{R},\downarrow}  
           \right)
          \left( c^{\dagger}_{\vec{R},\uparrow} c_{\vec{R},\uparrow} -
           c^{\dagger}_{\vec{R},\downarrow} c_{\vec{R},\downarrow} \right) 
\nonumber \\
  & &- \frac{J}{4} \sum_{\vec{R}}
 \left(   f^{\dagger}_{\vec{R},\downarrow} c_{\vec{R},\downarrow}
               + c^{\dagger}_{\vec{R},\uparrow} f_{\vec{R},\uparrow}  
       \right)^2
      +
         \left(  f^{\dagger}_{\vec{R},\uparrow} c_{\vec{R},\uparrow}
             + c^{\dagger}_{\vec{R},\downarrow} f_{\vec{R},\downarrow}  
\right)^2 \nonumber
\end{eqnarray}
where the last term accounts precisely for the spin flip terms:
$ \frac{J}{2} \sum_{\vec{R}}
 \left( f^{\dagger}_{\vec{R},\uparrow} f_{\vec{R},\downarrow}
 c^{\dagger}_{\vec{R},\downarrow} c_{\vec{R},\uparrow}  +
 f^{\dagger}_{\vec{R},\downarrow} f_{\vec{R},\uparrow}
 c^{\dagger}_{\vec{R},\uparrow} c_{\vec{R},\downarrow} \right) $.
With the above rewriting of the KLM  we can introduce order parameters 
both for magnetism and Kondo screening.
\begin{eqnarray}
& &
\frac{1}{2} \langle f^{\dagger}_{\vec{R},\uparrow} f_{\vec{R},\uparrow} -
               f^{\dagger}_{\vec{R},\downarrow} f_{\vec{R},\downarrow}  \rangle =
       m_f e^{i \vec{Q} \cdot  \vec{R} }  \nonumber \\
& &   \frac{1}{2} \langle c^{\dagger}_{\vec{R},\uparrow} c_{\vec{R},\uparrow} -
                  c^{\dagger}_{\vec{R},\downarrow} c_{\vec{R},\downarrow}  \rangle =
      - m_c e^{i \vec{Q} \cdot \vec{R} } \; \; \;  {\rm and}  \nonumber \\
& &  \langle f^{\dagger}_{\vec{R},\downarrow} c_{\vec{R},\downarrow}
        + c^{\dagger}_{\vec{R},\uparrow} f_{\vec{R},\uparrow} \rangle
   = \langle  f^{\dagger}_{\vec{R},\uparrow} c_{\vec{R},\uparrow}
        + c^{\dagger}_{\vec{R},\downarrow} f_{\vec{R},\downarrow} \rangle = -V. 
\end{eqnarray}
where $\vec{Q}$ correspond to the relevant wave vector for magnetic order.
Concentrating in a first step on paramagnetic solutions $m_s = m_f =0 $ we 
obtain the mean-field Hamiltonian
\begin{eqnarray}
    H &=& -t \sum_{\langle \vec{i},\vec{j} \rangle ,\sigma} 
      \left( c^{\dagger}_{\vec{i},\sigma} c_{\vec{j},\sigma}   + {\rm H.c.} \right)
    - \mu  \sum_{\vec{i},\sigma} c^{\dagger}_{\vec{i},\sigma} c_{\vec{i},\sigma }
    \nonumber \\
   & & +  \frac{J V  }{2} \sum_{\vec{R},\sigma} 
 \left( f^{\dagger}_{\vec{R},\sigma}c_{\vec{R},\sigma}   + {\rm H.c.} \right) 
   \nonumber \\
    & & +    \lambda  \sum_{\vec{R}} \left( \sum_{\sigma} 
       f^{\dagger}_{\vec{R},\sigma } f_{\vec{R},\sigma } - 1 \right) + 
 \frac{JN_{imp} V  ^2}{2}
\end{eqnarray} 
where $\mu $ corresponds to the chemical potential, $\lambda$ to a Lagrange 
multiplier enforcing the constraint of single occupancy on the $f$-sites  on
average and $N_{imp}$ is the number of impurity spins. The saddle point equations, 
$\frac{ \partial F }{\partial  \lambda  } 
 = \frac{ \partial F } { \partial  V   } = 0, \; \;
\frac{1}{L^2} \frac{ \partial F } {\partial  \mu  } = n_c $ where $F$ is the
free energy may then be 
solved to obtain the mean-field solution.   The thus obtained saddle point
corresponds precisely  to that obtained in the large-N approximation 
\cite{Newns87,Hewson,Georges00}. 
\begin{figure}
\noindent
\includegraphics[width=0.23\textwidth,height=0.15\textheight]{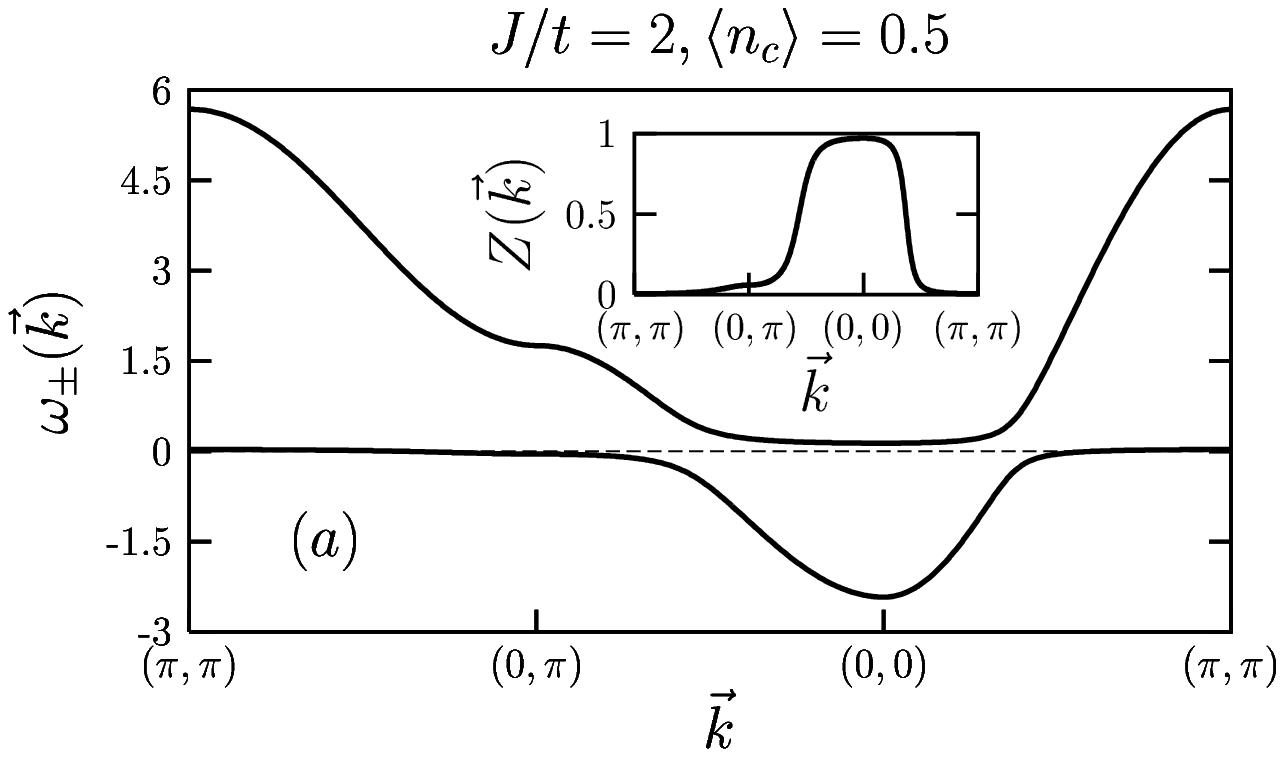} \hspace*{0.2cm}
\includegraphics[width=0.23\textwidth,height=0.15\textheight]{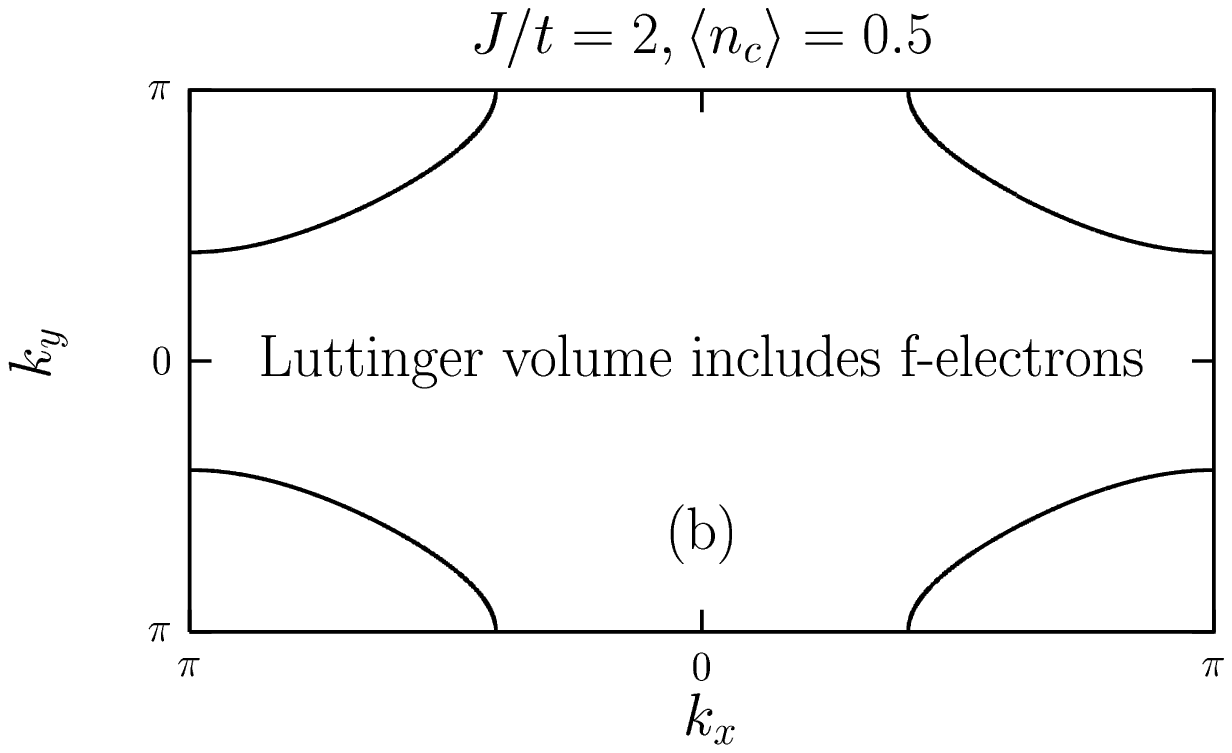} \\
\includegraphics[width=0.23\textwidth,height=0.15\textheight]{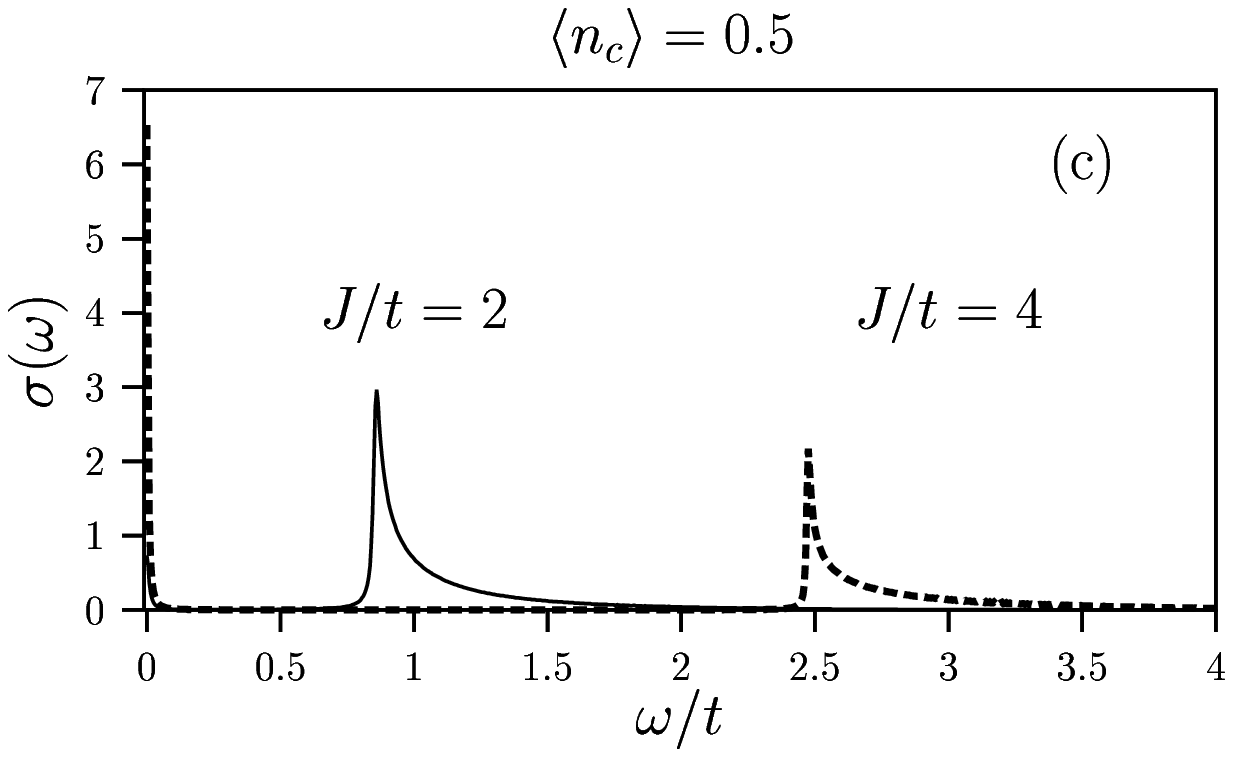} \hspace*{0.22cm}
\includegraphics[width=0.23\textwidth,height=0.15 \textheight]{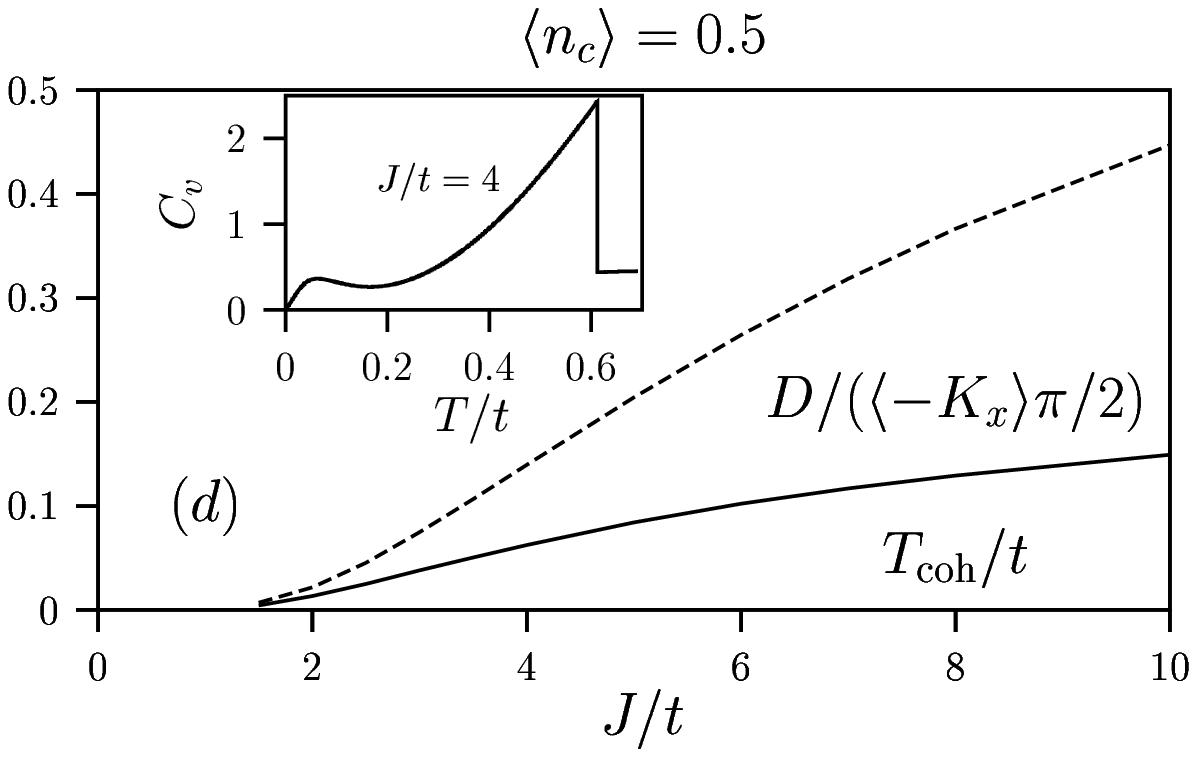}
\caption[]{Paramagnetic mean-field solution of the KLM at $n_c = 0.5 $ (a) The band structure. 
Energies are measured with respect to the Fermi energy. The inset corresponds to the
quasiparticle weight of the conduction band.  (b) Fermi line. (c) Optical 
conductivity. Note that there is a Drude peak at $\omega = 0$. (d) Drude weight 
normalized by the sum rule as  well as the coherence temperature as a function 
of $J/t$.  The inset corresponds to the specific heat from which we obtain the 
coherence temperature. (see text)
}
\label{KLM_PM}
\end{figure}

Fig. \ref{KLM_PM} summarizes the 
paramagnetic mean-field solution for the KLM at $n_c = 0.5$. The band structure 
plotted in Fig. \ref{KLM_PM}a consists of two bands (hybridized bands). 
As apparent, the band at the Fermi energy is flat and has a small quasiparticle 
weight (Fig. \ref{KLM_PM}a inset). The corresponding Fermi surface  
(Fig. \ref{KLM_PM}b),  covers $75\%$ of the Brillouin zone, so that both 
$f-$electrons and conduction electrons are included in the Luttinger volume. 
Given the band structure, the optical conductivity follows (Fig. \ref{KLM_PM}c).
The large mass  
(i.e. combined  small quasiparticle weight and flat dispersion relation ) 
of the  conduction electrons at the Fermi energy leads to a small Drude weight. 
The feature at higher frequencies are nothing but interband transitions.
The Drude weight normalized by the sum rule as a function of $J/t$ 
is plotted in Fig. \ref{KLM_PM}d. At small values of $J/t$ it becomes 
arbitrarily small reflecting the large mass of the charge carries. Along with 
the small Drude weight one expects a small coherence temperature. The coherence 
temperature ( $T_{\rm coh}$ ) - the energy scale at which the Fermi liquid
character of the ground state becomes apparent - may be defined for example 
by the energy scale below which the specific heat is linear in $T$ (inset
Fig. \ref{KLM_PM}d). As expected the $J/t$ dependence of $T_{ \rm coh}$ tracks
the Drude weight. 
In the mean-field approach the Kondo temperature, $T_K$,
corresponds to
the temperature scale at which the order parameter $V$ vanishes. Below $T_K$ the 
magnetic impurities are screened. In the mean-field approximation, this 
screening does not take place via the formation of a many-body 
state  of the impurity spin and conduction electrons but via charge
fluctuations on the impurity sites. In this approximation, charge 
fluctuations are  allowed 
since the constraint of single occupancy of $f$-sites is only imposed
on average.
As apparent by the jump in the specific heat (see inset of Fig. \ref{KLM_PM}d) 
$T_K$ corresponds to a a phase transition  rather 
than to a crossover scale. For a recent discussion of the Kondo and coherence 
scales within this approximation, the reader is referred to \cite{Georges00}.

\begin{figure}
\noindent
\begin{center}
\includegraphics[width=0.22\textwidth,height=0.15\textheight]{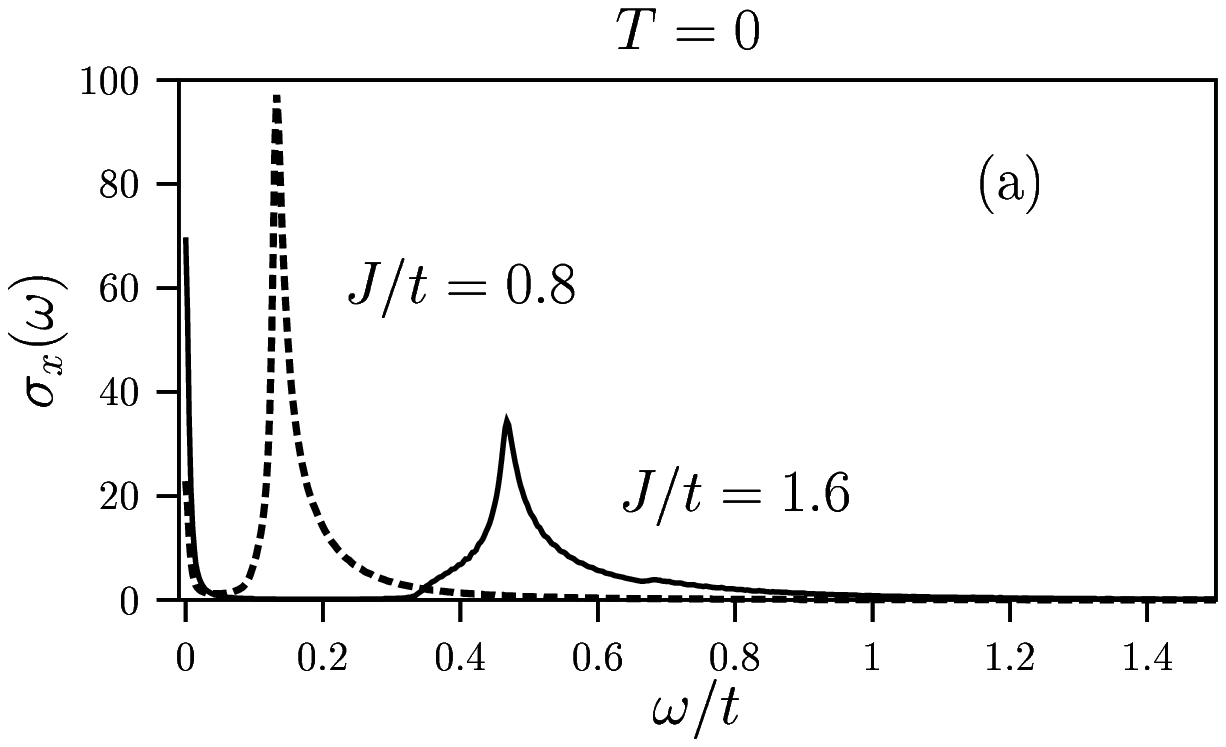} \hspace*{0.4cm}
\includegraphics[width=0.22\textwidth,height=0.15\textheight]{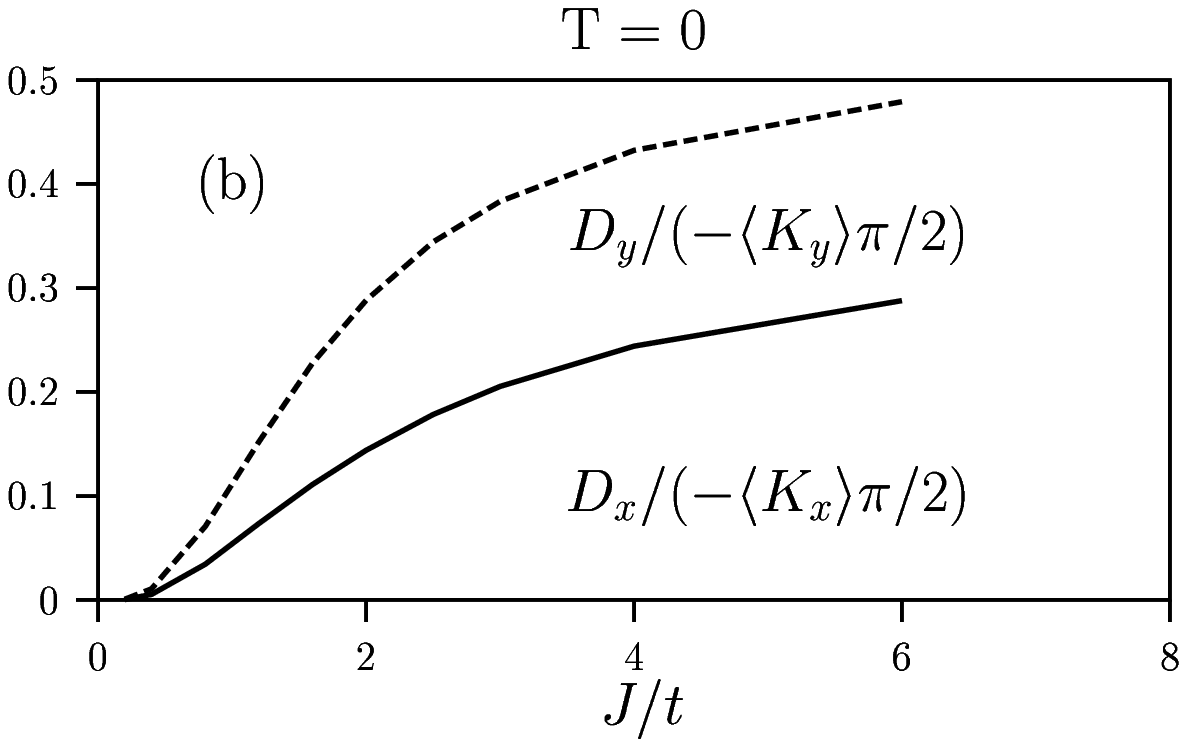} \\
\includegraphics[width=0.22\textwidth,height=0.12\textheight]{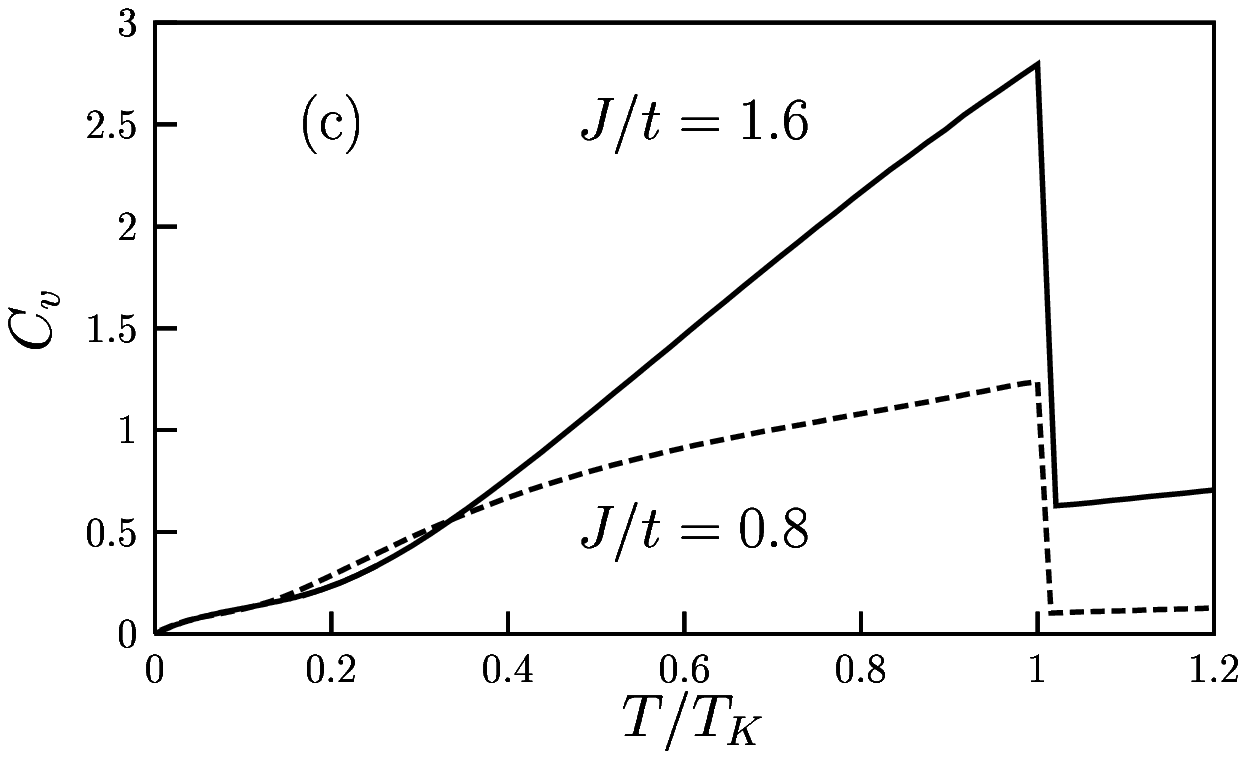}
\end{center}
\caption[]{ Paramagnetic mean-field  solution for the DKLM at $n_c = 4$.  ( 
i.e. the unit cell contains four  conduction electrons) (a) Optical conductivity.
(b) Drude weight normalized by the sum rule. (c) Specific heat versus $T/T_K$. 
}
\label{DKLM_PM}
\end{figure}
We now turn our attention to the DKLM. Figs. \ref{DKLM_PM}a,b plot the optical
conductivity as well as the Drude weight versus $J/t$ within the paramagnetic
mean-field approximation. As apparent, there is a clear similarity with the 
above presented results for the KLM. From the specific heat data, 
(Fig.  \ref{DKLM_PM} c) one can extract the coherence temperature. Here, some
care has to be taken since the model has a van-Hove  singularity at the 
Fermi level. Hence the coherence temperature  corresponds to the energy 
scale below which the specific heat follows a $ T \ln (1/T) $ law. 
Hence $T_{\rm coh} \sim 0.1 T_K$ for the considered couplings.

\begin{figure}
\includegraphics[width=0.45\textwidth]{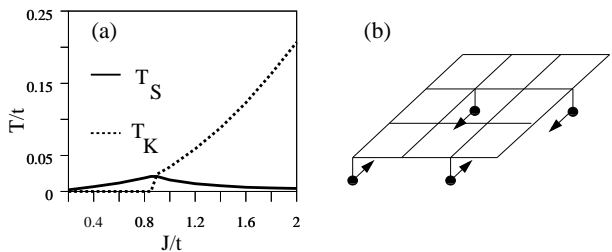}
\caption[]{(a) Mean-field energy scales for the DKLM at $n_c = 4$.  $T_K$ 
corresponds to the Kondo scale below which $V \neq 0$. $T_S$ corresponds to 
a magnetic energy scale below which magnetic ordering  with wave vector 
$\vec{Q} = \vec{b}_1/2$ --  schematically shown in Fig. (b) -- occurs.
$T_K$ and $T_S$ mark phase transitions. For $J/t > 0.8$ both Kondo 
screening ($ V \neq 0$) and magnetic ordering ($m_s \neq 0$, $m_f \neq 0$) 
are present in the ground state.  At $J/t < 0.8$ the Kondo scale  vanishes
since in the mean-field approximation  magnetic ordering freezes the 
impurity spins. The coherence temperature  (not plotted in the 
figure) lies roughly an order of magnitude lower that the Kondo scale. 
Hence  for most of the considered couplings it will be  masked by the magnetic 
instability.
}
\label{Scales_MF.fig}
\end{figure}

Next, we  consider magnetic degrees of freedom for the DKLM.  
As mentioned above, at
$J/t = \infty$  the model maps onto free electrons. The underlying particle-hole
symmetry of the model directly leads to nesting so that the $J/t = \infty$ point 
will be unstable to magnetic ordering with wave vector $\vec{b}_1/2$
(Fig. \ref{latt.fig}c).
At weak  couplings  the Ruderman-Kittel-Kasuya-Yossida
(RKKY) interaction
 favors  magnetic order rather that Kondo screening \cite{Doniach77}.
Hence,  for all values of $J/t$ we anticipate  a 
magnetically ordered ground state. 
Fig. \ref{Scales_MF.fig} plots the mean-field transition temperatures  
corresponding to Kondo screening, $T_K$, (below $T_K$, $V \neq 0$) 
and  magnetism, $T_S$
($ m_s \neq 0$ $m_f \neq 0$).
As apparent, magnetic order is always present.
For values of $J/t > 0.8$ the mean-field
Kondo scale exceeds the magnetic scale $T_S$.
For this parameter range 
and as a function of decreasing temperature, one expects to be able to study
the formation of the heavy electron state prior to the onset of magnetic
fluctuations which due to nesting will drive the system to an insulator. 
The QMC simulations  presented in section \ref{QMC_res}  are aimed at studying
this crossover.

\section{Numerical techniques}
In this section we will first briefly summarize the numerical techniques
used. We then turn our attention  on a simple method to dramatically reduce the 
size effects for the simple case of free electrons. This  is the 
key for producing size independent results down to low temperatures for the
DKLM. However, this method  is general and may  prove  useful for 
other applications.

\subsection{Quantum Monte Carlo algorithms}
We have used ground state and finite temperature versions of the 
auxiliary field QMC method to investigate the DKLM.  The $T=0$ method is
based on the equation:
\begin{equation}
  \frac{\langle \Psi_0 | O | \Psi_0 \rangle }
       {\langle \Psi_0 | \Psi_0 \rangle }   =
\lim_{\Theta t \rightarrow \infty}
   \frac{ \langle \Psi_T | e^{-\Theta H_{DKLM}} O  e^{-\Theta H_{DKLM} 
       | \Psi_T \rangle } }
       {\langle \Psi_T | e^{-2\Theta H_{DKLM}}  | \Psi_T \rangle }
\end{equation}
where $ |\Psi_T \rangle $ is required to be  non-orthogonal to the ground 
state $ |\Psi_0 \rangle $.
Finite temperature properties are calculated in the grand canonical 
ensemble via:
\begin{equation}
\langle  O  \rangle  = 
  \frac{ {\rm Tr } \left[  e^{-\beta H_{DPKLM}} O \right] }
       { {\rm Tr } \left[  e^{-\beta H_{DPKLM}} \right]  }.
\end{equation}
Details of how to implement the QMC method without generating a sign
problem in the particle-hole filled case have been introduced an extensively
described for the KLM in Refs \cite{Assaad99a,Capponi00}. Since precisely the
same {\it tricks} may be used for the DKLM, we refer the reader to those articles
for further details. We now concentrate on another important issue, size effects,
which become extremely severe when one wishes to study  metallic states 
down to low temperatures.  

Finally, to determine the single impurity Kondo scale for our given 
band-structure, we have used the Hirsch-Fye impurity algorithm \cite{HirschFye86}.
The same formulation as the lattice problem is readily implemented
in this approach.  

\subsection{Size effects and magnetic fields}
\label{SizeB.chap}
Reducing size effects in numerical simulations is important, since many
models of correlated electron systems may be solved numerically only on
{\it small} lattices. Size effects become particularly severe when the 
ground state turns out be be a metallic state with large coherence 
temperature. On the other hand, insulators are characterized by the 
localization of the wave function and are hence rather insensitive 
to boundary conditions on finite sized systems. It thus becomes apparent, 
that the worst case scenario for severe size effects are just free electrons
in  a tight binding approximation:
\begin{equation}
         H = -t \sum_{\langle \vec{i},\vec{j} \rangle }  
     c_{\vec{i}}^{\dagger}  c_{j} + {\rm H.c.}.
\end{equation}
In many cases before turning on the interaction  which will automatically
restrict the size of the lattice under consideration it is important to 
control size effects for this simple case. 
We will concentrate on the two dimensional case on a torus geometry which 
for the above model reduces to imposing periodic boundary conditions:
$c_{\vec{i} + L \vec{e}_{x} }^{\dagger} = c^{\dagger}_{\vec{i}}$,
$c_{\vec{i} + L \vec{e}_{y} }^{\dagger} = c^{\dagger}_{\vec{i}}$
where  $L$ is the linear length of the lattice lying in the 
$\vec{e}_x, \vec{e}_y$ plane.

To reduce size effects on  thermodynamic quantities
one may in principle consider the Hamiltonian:
\begin{equation}
 H(L) =  \sum_{\langle \vec{i},\vec{j} \rangle }  t_{\vec{i},\vec{j}}(L) 
     c_{\vec{i}}^{\dagger}  c_{j} + {\rm H.c.}
\end{equation}
where $t_{\vec{i},\vec{j}}(L)$ are arbitrary hopping parameters which have to
satisfy 
\begin{equation}
\label{HopL}
  \lim_{L \rightarrow \infty } t_{\vec{i},\vec{j}}(L)    = -t.
\end{equation}
Clearly this choice of hopping matrix elements on finite lattices will 
break the lattice symmetry. This is  a price we are willing to pay  provided 
that the convergence as a function of system size 
of thermodynamics quantities is greatly improved. 
Eq. \ref{HopL} nevertheless guarantees that  in  the thermodynamic limit 
this symmetry is restored. 
To determine the hopping matrix elements $ t_{\vec{i},\vec{j}}(L) $ so 
as to reduce  size effects on say the specific heat, $ C_v(L,T) = 
\frac{\partial E(L)}{ \partial T} $,  one may minimize
\begin{equation}
      \chi^2 = \sum_{T} \left[ C_v(L,T) - C_v(L = \infty,T)  \right]^2
\end{equation}   
where the sum extends over a given range of temperatures. Taking into account 
only amplitude modulations of the hopping matrix elements leads already 
to a cumbersome minimization problem which does not provide satisfactory results.

Instead of carrying out a complicated minimization problem we can try 
to guess which matrix elements $t_{\vec{i},\vec{j}}(L)$ will minimize 
size effects. It turns out that introducing a magnetic field produces
remarkable results. The magnetic field is introduced via the Peirls phase
factors:
\begin{equation}
 H(L) = -t\sum_{\langle \vec{i},\vec{j} \rangle } 
        e^{\frac{2 \pi i}{\Phi_0} \int_{\vec{i}}^{\vec{j}} \vec{A}_L (\vec{l})
  \cdot {\rm d} \vec{l} }
     c_{\vec{i}}^{\dagger}  c_{j} + {\rm H.c.}
\end{equation}
with
$ \vec{B}_L(\vec{x}) = \vec{\nabla} \times  \vec{A}_L (\vec{x})$ and
$\Phi_0$ the flux quanta. 
The torus geometry imposes restrictions on the $\vec{B}_L $ field. 
Since, a translation in the argument of the vector potential may 
be absorbed in a gauge transformation:
\begin{eqnarray}
   \vec{A}_L(\vec{x} + L\vec{e}_{x} ) & = & \vec{A}_L(\vec{x}) +
   \vec{\nabla} \chi_{x} (\vec{x}), \nonumber \\
   \vec{A}_L(\vec{x} + L\vec{e}_{y} ) & = & \vec{A}_L(\vec{x}) +
   \vec{\nabla} \chi_{y} (\vec{x}), \; \;
\end{eqnarray}
we  chose, the boundary condition 
\begin{equation}
c_{\vec{i} + L\vec{e}_{x} }^{\dagger} =
 e^{ \frac{2\pi i}{\Phi_0} \chi_{x} (\vec{i}) }
  c_{\vec{i}}^{\dagger},         \; \;
c_{\vec{i} + L\vec{e}_{y} }^{\dagger} =
 e^{ \frac{2\pi i}{\Phi_0} \chi_{y} (\vec{i}) }
  c_{\vec{i}}^{\dagger} 
\end{equation}
to satisfy the requirement:
\begin{equation}
  [H(L),T_{L \vec{e}_x} ] = [H(L),T_{L \vec{e}_y} ] = 0.
\end{equation}
Here, $T_{\vec{x}}$ corresponds to a translation by $\vec{x}$. 
However, magnetic translation operators belong to the magnetic 
algebra \cite{Fradkin91}:
\begin{equation}
       T_{L \vec{e}_x} T_{L \vec{e}_y} = 
e^{-i 2 \pi \frac{(L\vec{e_x} \times L\vec{e}_y) \cdot \vec{B}}{\Phi_0} } 
        T_{L \vec{e}_y} T_{L \vec{e}_x}.
\end{equation}
Thus, to obtain a single valued    
wave function the condition of flux quantization has to 
be satisfied: 
$\frac{(L\vec{e_x} \times L\vec{e}_y) \cdot \vec{B}}{\Phi_0 } = n$
where $n$ is an integer. 
Here, we consider a static magnetic field running along the $z$-axis 
perpendicular to the $x,y$ plane in which the lattice lies. Hence, the 
smallest magnetic field  which we can consider on a given lattice size 
satisfies:
\begin{equation}
        \frac{B  L^2}{\Phi_0} =  1.
\end{equation}
With this choice of magnetic field  and associated vector potential 
Eq. \ref{HopL} holds.

\begin{figure}
\includegraphics[width=0.45\textwidth]{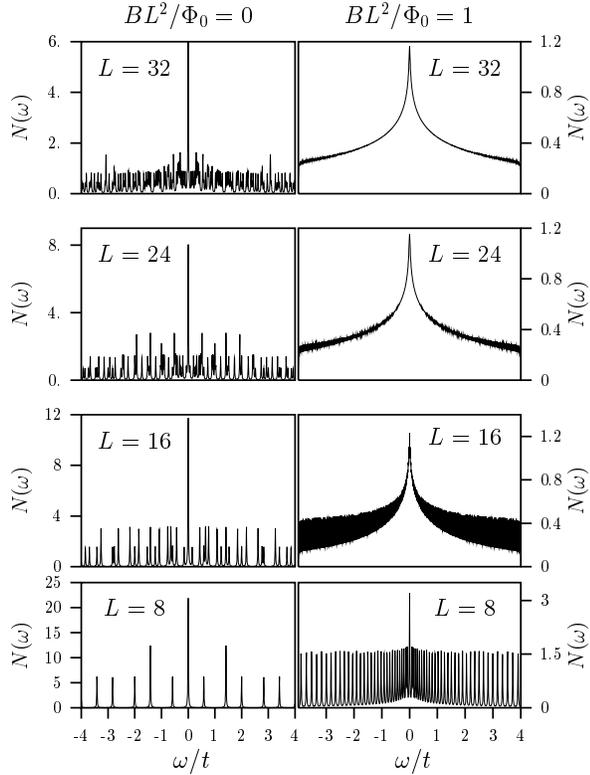}
\caption[]{ Density of states $N(\omega) = \frac{1}{N}\sum_{r} N(r,\omega) $
with (right column) and without (left column)  magnetic field. Here, 
we consider $\delta = 0.01 t$.
}
\label{Nom.fig}
\end{figure}
To illustrate the reduction of size effects caused by the inclusion of the 
magnetic field, we first consider the single particle density of states.
In a basis where $H(L)$ is diagonal,  $ H(L) = \sum_{n= 1}^{N} \epsilon_n 
\gamma^{\dagger}_{n}\gamma_n$ with  $ c^{\dagger}_n = \sum_{m} 
\gamma^{\dagger}_m U^{\dagger}_{m,n} $ and $U^{\dagger} U = I$, the local
density of states reads:
\begin{equation}        
        N(r,\omega) = {\rm Im} \sum_{n}^{N} \frac{|U_{n,r} |^2}
        { \epsilon_n - \omega - i \delta}
\end{equation}
where $\delta $ is a positive infinitesimal and $N$ the total number of sites.
Since the magnetic field breaks translation invariance (it is the site 
dependent vector potential which enters the Hamiltonian) $N(r,\omega)$ is site
dependent. Averaging over sites yields the density of states 
$N(\omega)$ plotted in Fig. \ref{Nom.fig}.  As apparent,  without the 
magnetic field
and up to $L=32$, $N(\omega)$ is dominated by size effects for the considered
value of $\delta = 0.01 t$. In contrast, the presence of the magnetic 
field provides remarkable improvements. In particular the van-Hove 
singularity is well reproduced already on $ L = 16$ lattices and at $L=32$
the result is next to exact for the considered value of $\delta$.
It is instructive to look at 
the $L=8$ case with and without magnetic fields. When $B$ is turned  on,  
the degeneracy of levels is lifted. Each level - apart from the 
$\epsilon_n = 0$ level which is two fold degenerate - is nondegenerate.
This is precisely what one expects for Landau levels which have degeneracy 
$L^2 B /\Phi_0$ which is unity in our case. This provides an intuitive 
understanding of why the method works so well. Since each level becomes 
singly degenerate, the single particle states cover homogeneously the 
the energy range of the band-width. Clearly this can only be achieved
by breaking the lattice symmetry on finite sized systems. 

\begin{figure}
\begin{center}
\includegraphics[width=0.45\textwidth,height=0.15\textheight]{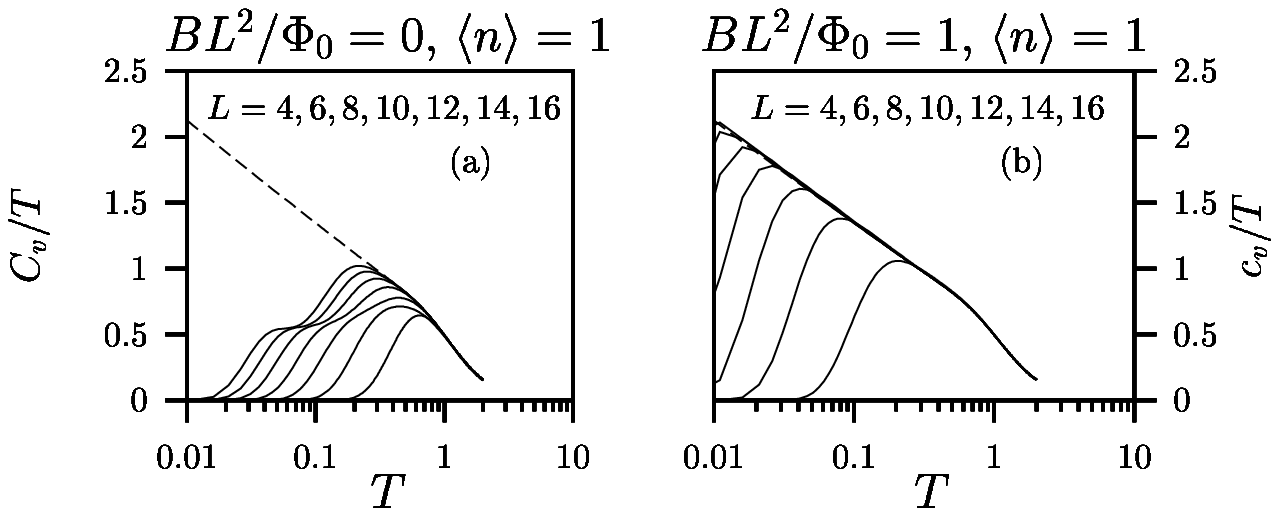} \\
\includegraphics[width=0.45\textwidth,height=0.15\textheight]{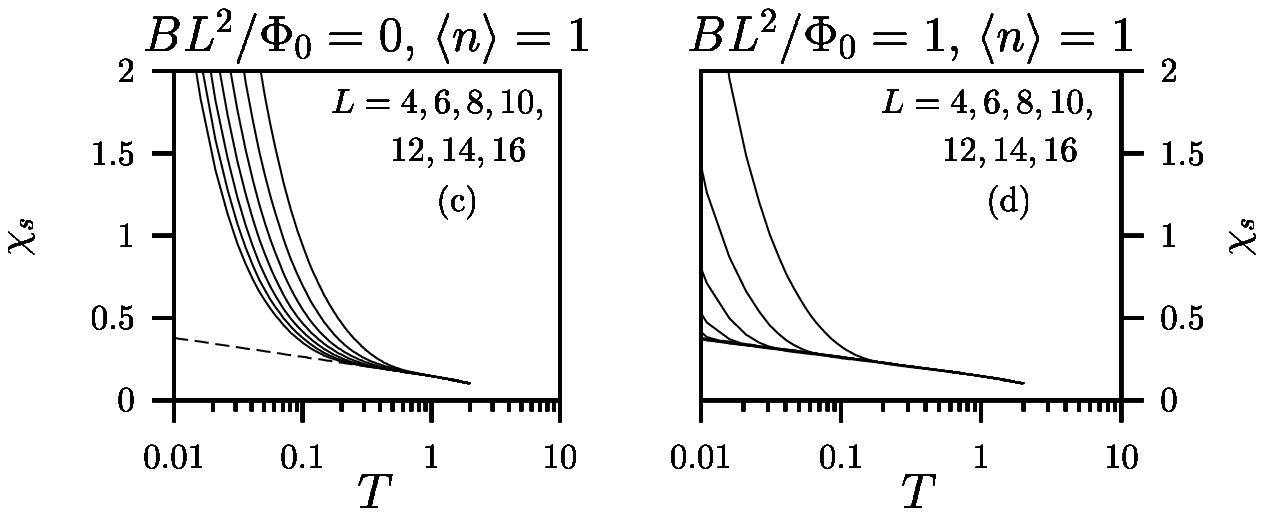}
\end{center}
\caption[]{Specific and spin susceptibility versus temperature  without  (a,c) 
and with (b,d) magnetic field.  
The curves from right to left correspond to  increasingly
large lattices as denoted in the figure. The dashed line corresponds to the
exact result. 
}
\label{Cv.fig}
\end{figure}

We now turn our attention to the specific heat coefficient $\gamma = C_v/T$
(see Fig. \ref{Cv.fig}a,b). As apparent, for a given system size, 
the inclusion of the magnetic field provides a 
gain of more than one order of magnitude in the
temperature scale at which size effects set in.  In particular the 
$\ln (1/T) $ behavior of $\gamma$ due the the van-Hove singularity  becomes 
apparent already on $L=6$ lattices. 

Upon inspection a similar improvement is obtained for the spin
susceptibility (see Fig. \ref{Cv.fig}c,d). Note 
that since we are dealing with free electrons the charge and spin  
susceptibilities  are identical.

 
One crucial question is whether the magnetic field will introduce a sign 
problem in the numerical simulations. It turns out that for  DKLM at the 
particle-hole symmetric point it does not \cite{Capponi00}.
Hence we can incorporate 
it in our simulations
and benefit from the above demonstrated drastic improvement in reduction
of size effects. 

Other schemes have been proposed to reduce size effects. In particular, 
averaging over boundary conditions has been suggested \cite{Poilblanc91,Gross92}.
This method
has the advantage of not breaking translation symmetry. However, the averaging
requires several simulations and is hence computationally expensive. In contrast
with the presented method the improvement in reduction
of size effects is obtained within a single simulation.

\section{Numerical Results}
\label{QMC_res}
Our numerical results for the DKLM are summarized in this section. 
We first concentrate on ground-state properties and 
check for magnetic ordering as a function of $J/t$.
Various thermodynamic quantities (specific heat, spin and charge 
susceptibilities) 
as well as the optical conductivity and hence resistivity are then 
analyzed as a function of temperature at different values of $J/t$.

\subsection{Spin-spin correlations at T= 0}
To detect long-range magnetic ordering, we compute the spin-spin correlation
function at $T=0$:
\begin{equation}
   S(\vec{Q}) = \sum_{\vec{R}} e^{i \vec{R} \cdot \vec{Q} } \frac{4}{3} 
                \langle \vec{S}^{f}_{\vec{R} = 0} \cdot 
                        \vec{S}^{f}_{\vec{R}} \rangle .
\end{equation}
Here we consider $\vec{Q} = \vec{b}_1/2$ and the sum runs over the impurity
spins. In the presence of long-range magnetic order, $ S(\vec{Q}) $ is an 
extensive quantity so that the staggered moment, 
$ m_s = \lim_{L \rightarrow \infty}\sqrt{ S(\vec{Q})/L^2 }$, is  finite.
$L^2$ denotes the number of unit-cells. \\
\begin{figure}
\includegraphics[width=0.45\textwidth]{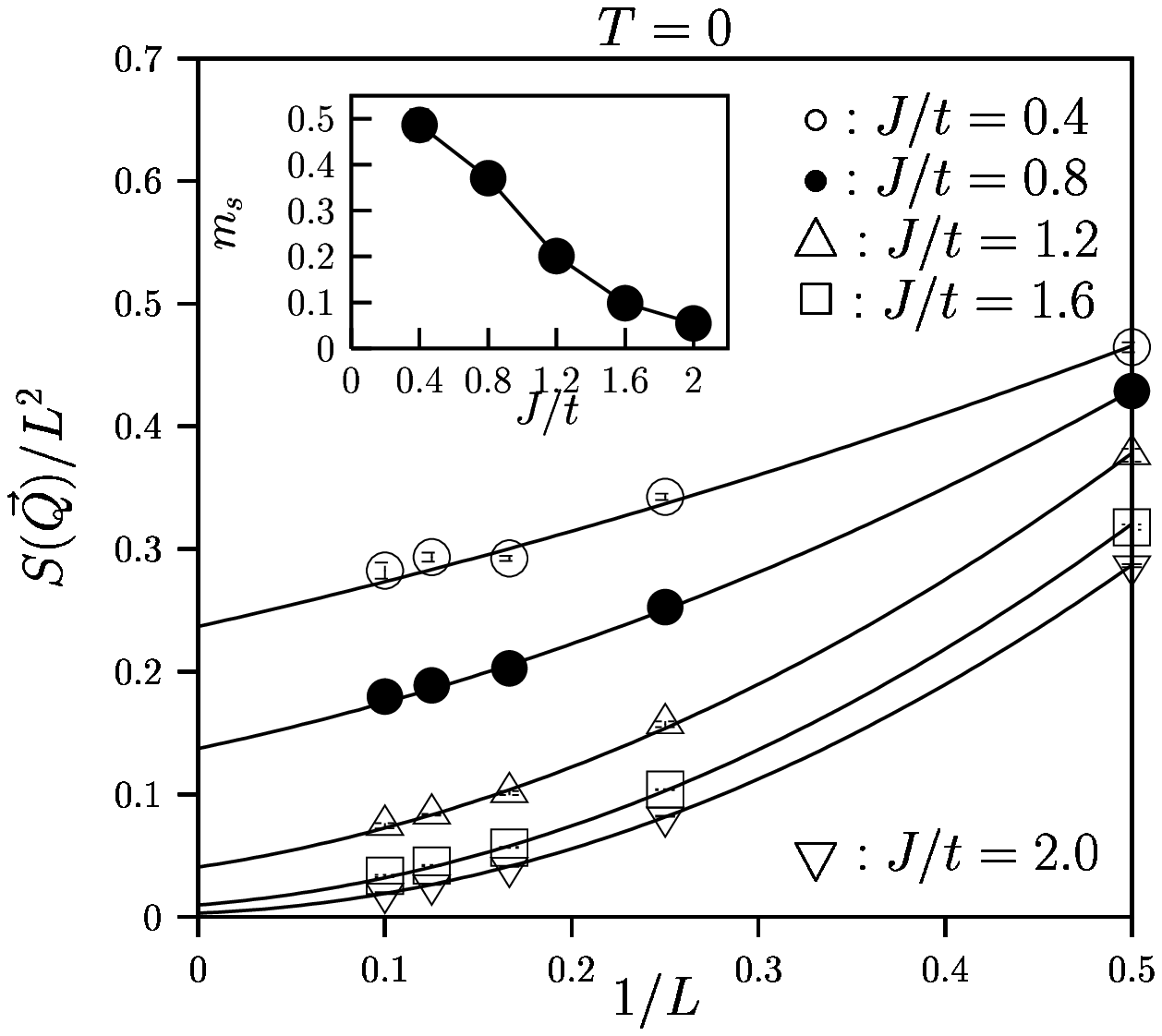}
\caption[]{  $S(\vec{Q})$ at $T=0$ for the DKLM
 as a function of system
size and  various values of $J/t$. The $L=10$ point corresponds to a lattice
of $10 \times 10$ unit cells and hence 500 orbitals. After extrapolation 
to the thermodynamic limit, we obtain the staggered moment plotted in the inset.
}
\label{SpinT0.fig}
\end{figure} 
Figure \ref{SpinT0.fig} plots $S(\vec{Q})$ for various values of $J/t$. 
The inset in Fig. \ref{SpinT0.fig} plots the staggered moment obtained after
extrapolation of $S(\vec{Q})$ to the thermodynamic limit. The data is thus 
consistent with long-range magnetic ordering at least up to $J/t = 2$. Clearly
as $J/t$ grows and the staggered moment becomes small, it is increasingly hard to
distinguish it from zero. Nevertheless, and based on the mean-field result, 
we expect the staggered moment to vanish only in the limit 
$J/t \rightarrow \infty$.

Following arguments put forward in Ref. \cite{Capponi00}   Kondo
screening and magnetic ordering to coexist: impurities are partially screened
and the remnant magnetic moment orders. Depending upon the relative energy gains
of both mechanisms a staggered moment of  arbitrary magnitude  can be produced.
In the weak coupling limit, the energy gain for magnetism is the RKKY interaction.
This interaction leads to an effective Heisenberg model between impurity
spins with  exchange $ J_{RKKY} (\vec{q}) = -J^2 {\rm Re}
\chi \left( \vec{q}, \omega = 0 \right) $ where 
$\chi \left( \vec{q}, \omega \right) $ 
corresponds to the spin susceptibility of the conduction electrons. 
On the other hand screening  occurs on an energy scale set by the Kondo
temperature: $T_K \sim W e^{-W/J} $ where $W $ is the band width. 
Hence  for small values of $J/t$ the system will gain energy by ordering 
magnetically. 
Although the wave vector of magnetic ordering will depend on the details
of the spin susceptibility, this source of energy gain is present  irrespective
of the band structure.
On the other hand, the strong coupling result is an artifact of the model.
In this limit, if
Kondo screening were complete, a nested Fermi surface would be produced as in
the $J/t \rightarrow \infty$ limit. Hence, in the ground state,  
the system will always gain energy by keeping Kondo screening partial. The remnant
magnetic moment will order with the nesting wave vector thereby  opening a 
gap and gaining energy.

Were it not for 
nesting a finite critical value of the coupling $J/t$ at
which a quantum phase transition between ordered and disordered magnetic 
states  would occur. Unfortunately, this is beyond the scope of the QMC  since
the absence of particle-hole symmetry and hence nesting leads to a sign problem.

\subsection{Transport and thermodynamic   properties.}
Here, we consider finite temperature properties of the model. As 
discussed above, the ground state is a magnetic 
insulator irrespective of $J/t$. The salient features of the model however, lie
in its thermodynamic properties. 

We first consider the real part of the  optical
conductivity as obtained from the Kubo formula, $\sigma(\omega,T)$. 
This quantity
is related to the imaginary time current-current correlation functions via:
\begin{eqnarray}
& & \langle J(\tau)J(0) \rangle = \int d \omega K(\omega,\tau) \sigma(\omega,T) 
\nonumber \\
& &  \; {\rm with} \; \; K(\omega,\tau) = \frac{1}{\pi}
\frac{e^{-\tau \omega} \omega}{1 - e^{- \beta \omega} }.
\label{kubo.eq}
\end{eqnarray}
Here $J$ is the current operator along  the $x$ or $y$ lattice direction and
$\langle \rangle$ represents an
average over the  finite-temperature ensemble. 
The above inverse Laplace transform,
to obtain the optical conductivity is carried out with the ME 
\cite{Jarrell96} method. We  use  the paramagnetic  mean-field solution 
with substantial broadening  
as default model. Note that the order parameter $V^2$ which is proportional
to the spin-spin correlation between the conduction electron and impurity spin
is estimated from the Monte Carlo run and not by solving the saddle point 
equations.
\begin{figure}[h]
\includegraphics[width=0.45\textwidth]{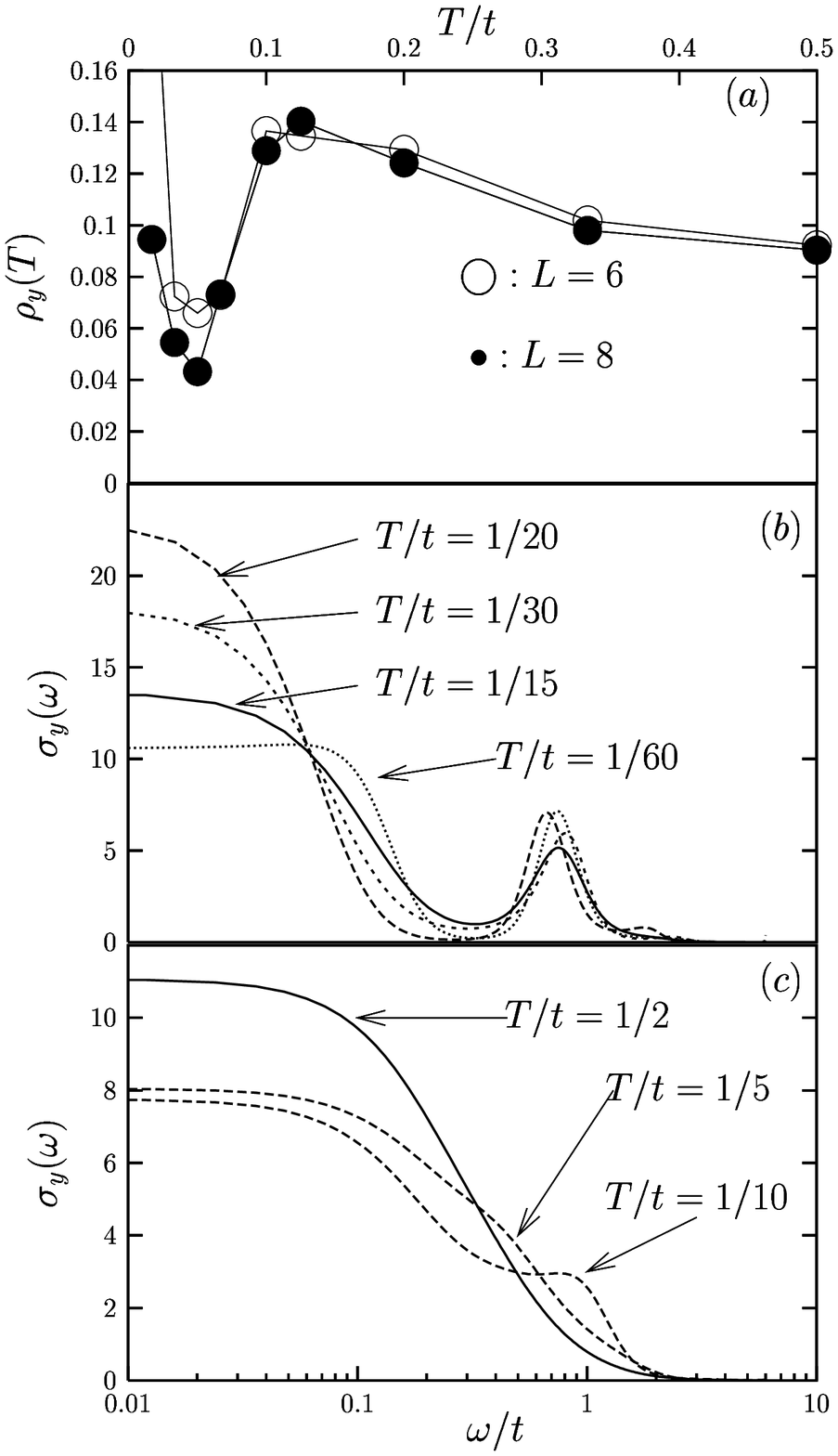}
\caption[]{ Optical conductivity and related resistivity  as a function of
temperature  at $J/t = 1.6$. The data in Figs. (b),(c) stem from simulations
on  lattices with $8 \times 8 $ unit cells (L=8).  In Fig. (a) both $L=6$ and
$L= 8$ results are included. As apparent and due to the inclusion of the 
magnetic field (see section \ref{SizeB.chap})
 size effects are next to absent until the magnetic length 
scale exceeds the lattice size (see text). }
\label{Cond16yy.fig}
\end{figure}
Our results for the real part of the optical conductivity and
associated resistivity  $\rho(T) = 1/\sigma(\omega = 0,T)$ are plotted in 
Fig. \ref{Cond16yy.fig}. Here we consider $J/t = 1.6$.
At high
temperatures, $T/t = 1/2$, the data consists of a single Drude feature. As the 
temperature is lowered to $T/t \sim 1/10$, the weight under the Drude peak is
reduced, and a feature at $\omega/ t \sim 1 $ is formed. A rise in the associated
resistivity as a function of decreasing temperature is observed in this temperature
range. This rise in resistivity originates from singular spin flip 
scattering on impurity spins and is reminiscent of the single impurity physics.
By further reducing the temperature to $T/t \sim 1/20 $ the Drude feature 
becomes sharper thus leading to a drop in the resistivity. Finally, at our 
lowest temperatures ($T/t = 1/30, 1/60$) there is a marked upturn in the 
resistivity.  At those temperatures the optical weight under the Drude peak
starts to shift to frequencies $\omega / t \sim 0.1$. 

\begin{figure}
\includegraphics[width=0.45\textwidth]{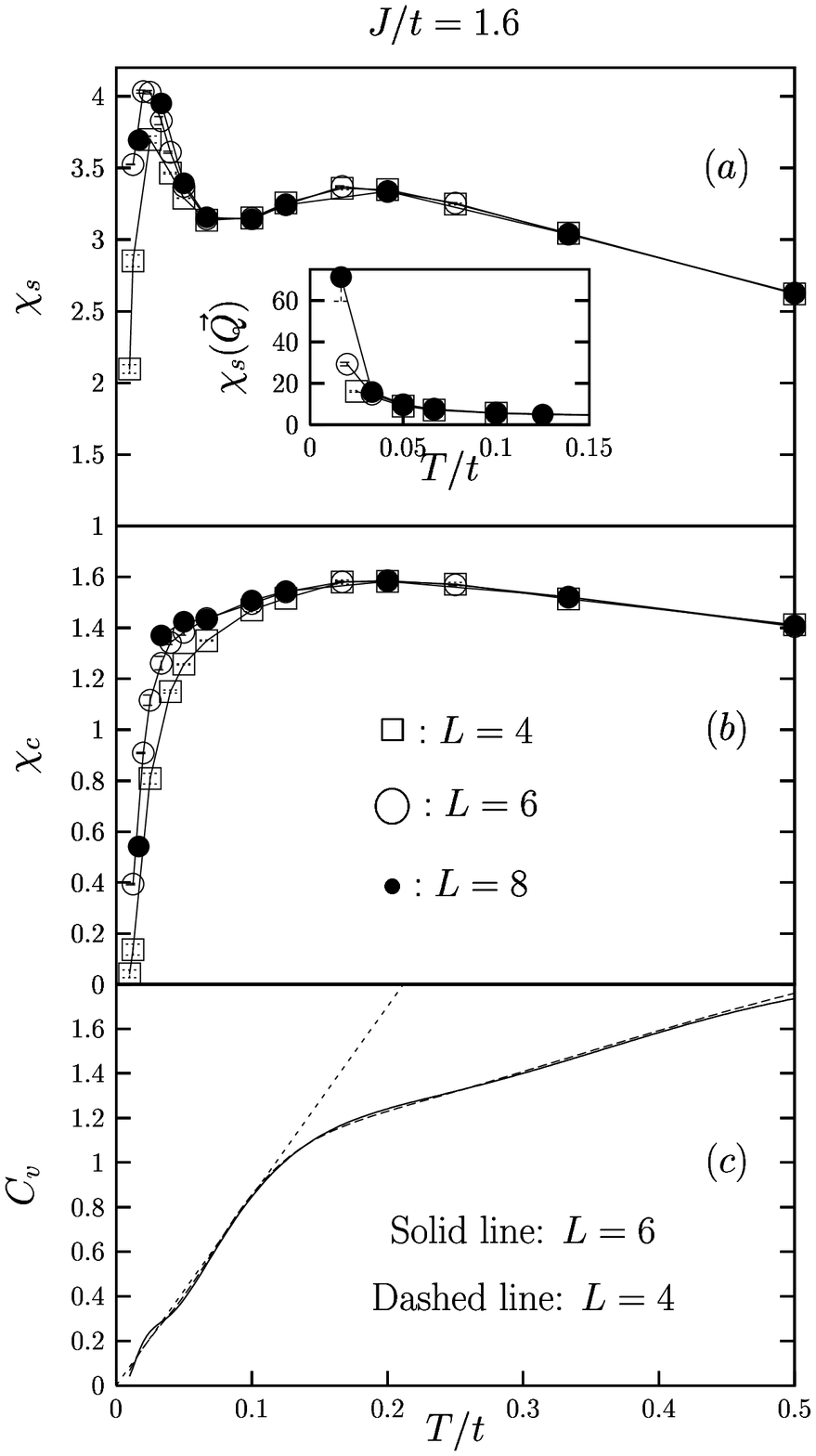}
\caption[]{  (a) Uniform and staggered (inset) spin susceptibilities (b)
charge susceptibilities and (c) specific heat for the DKLM at $J/t = 1.6$.
}
\label{Ther16.fig}
\end{figure}

The overall form of the resistivity versus temperature curve 
(Fig. \ref{Cond16yy.fig}a)  allows us to 
define three temperature scales. The highest energy scale we will  discuss 
corresponds to the resistivity minimum, $T_C$. For the considered coupling
$T_C \sim 0.5 t$. 
We will define $ T^{\star} \sim 0.1 t $ 
as the temperature scale at which a maximum in the resistivity is 
observed. The magnetic scale $T_S \sim 0.025 t$ is defined by the  energy 
scale below
which the $\rho(T)$ follows an {\it activated } behavior.  Next, we consider
thermodynamic properties and concentrate on their behavior 
at the above defined energy scales.

Fig. \ref{Ther16.fig} plots the specific heat $C_v$, charge and 
spin uniform susceptibilities ($ \chi_c$, $\chi_s$)
as well as the staggered spin susceptibility ($\chi_s(\vec{Q})$, 
$\vec{Q} = \vec{b}_1/2$)  as a function of temperature at 
$J/t = 1.6$. From the technical point of view, the specific heat 
is computed using the recently proposed ME based method of Ref. \cite{Huscroft00}.
As apparent in the temperature region $ T_S < T < T^{\star}$ one observes the 
following features. i)
As indicated in  Fig. \ref{Ther16.fig}c the specific heat is consistent
with a $\gamma T$  law.
ii) the uniform spin and charge susceptibilities after going through a 
broad maximum seem to saturate. Both points are the characteristics of 
a Fermi liquid.  
The staggered spin susceptibility essentially measures the  magnetic length
scale at the considered wave vector. The inset of Fig. \ref{Ther16.fig}a 
shows a marked increase in this quantity at the spin scale $T_S$. At the same energy
scale, a {\it sharp} peak in the spin susceptibility is observed. Following the 
size effects, the data is consistent with a saturation of this quantity in
the limit of zero temperature. 
At $T_S$ the charge susceptibility  \ref{Ther16.fig}b decreases rapidly.
Upon analysis of size effects the data is consistent with the vanishing of 
this quantity at $T=0$. Upon close analysis, an anomaly in the specific heat
is apparent at $T_S$.
Thus, $T_S$ marks the onset of substantial spin fluctuations which triggers
the formation of an insulating state. That is the vanishing of the charge 
susceptibility and an activated behavior of the resistivity below $T_S$.
 
\begin{figure}
\includegraphics[width=0.45\textwidth]{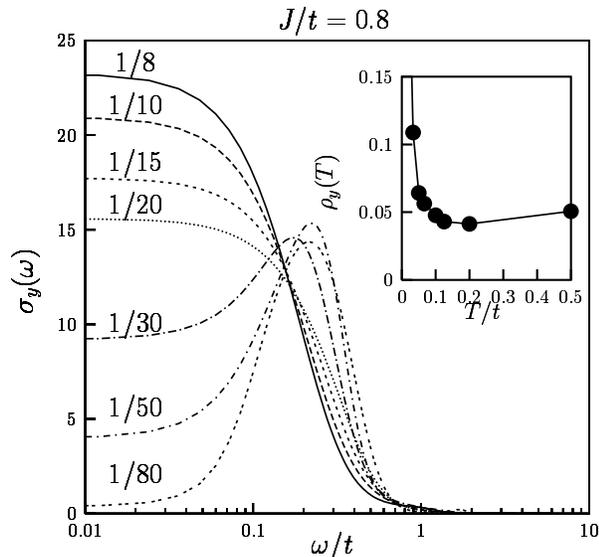}
\caption[]{ Optical conductivity as a function of temperature  at $J/t = 0.8$.
The inset  plots the resistivity versus temperature.  For those simulations
the lattice size is $L=6$.
}
\label{Cond08yy.fig}
\end{figure}
\begin{figure}
\includegraphics[width=0.45\textwidth]{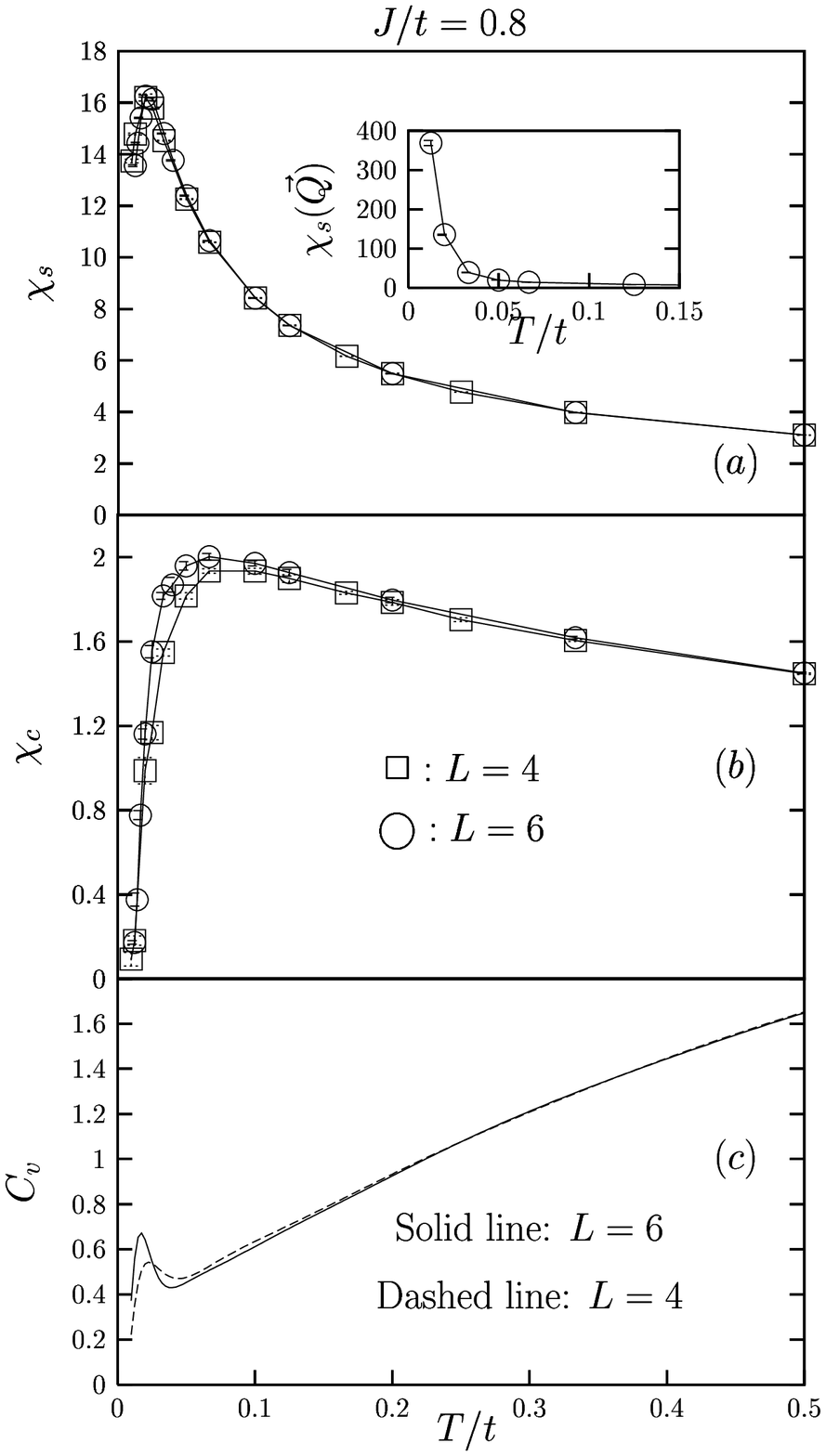}
\caption[]{Same as  Fig. \ref{Ther16.fig} but for $J/t = 0.8$.}
\label{Ther08.fig}
\end{figure}
Next, we consider  smaller couplings, $J/t = 0.8$.  Transport and 
thermodynamic properties are plotted in Figs.  \ref{Cond08yy.fig} and
\ref{Ther08.fig}.
As apparent from the conductivity data (Fig. \ref{Cond08yy.fig}), 
no maximum as a function of temperature 
is observed. After hitting a minimum at $T_C \sim 0.2t$ 
the resistivity grows monotonically as
a function of decreasing temperature. The thermodynamic data of Fig. 
\ref{Ther08.fig} equally shows no sign of the  $T^{\star}$ scale.
That is, there is no
energy scale for which the specific heat has a linear in $T$ dependence. The 
only scales which are apparent in Figs.\ref{Cond08yy.fig} and  \ref{Ther08.fig}
are resistivity minimum $T_C$ and the spin scale $T_S$.
At $T_S$ an anomaly in $C_v$ is detected, a drop in $\chi_c$ is observed and 
$\chi_s$ exhibits a sharp peak. The peak  in $\chi_s$ is again located at the
temperature scale at which $\chi_s(\vec{Q})$ shows a sharp increase.

\begin{figure}
\includegraphics[width=0.4\textwidth]{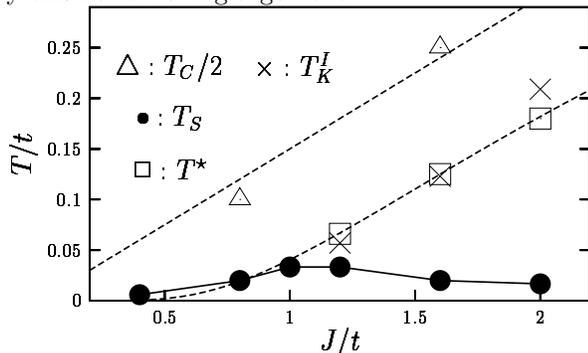}
\caption[]{ Crossover scales for the DKLM as a function of $J/t$. 
$T_C$ denotes the  resistivity minimum. $T^{\star} $ corresponds to a 
crossover scale at which a maximum in resistivity is observed.
Below $T_S$  at marked increase in
the magnetic length scale is observed.   We use the form $ e^{-1 / \alpha J} $ 
to fit the
$T^{star}$ and the form $ a T$ to fit the data for $T_C$. The crosses correspond
to the Kondo temperature of the single impurity problem. (See section \ref{TKI})
}
\label{Scales_qmc.fig}
\end{figure}

Our QMC results  are summarized in Fig. \ref{Scales_qmc.fig}. The highest
energy scale, $T_C$,  corresponds to the resistivity minimum, triggered 
by enhanced spin flip scattering off the impurity spins. 
As already observed 
for the KLM, this energy scale in a first approximation tracks $J$.
At high temperatures, it is tempting  to try to understand this energy 
scale in terms of the single impurity Kondo model. In this case, 
we can use the closed  form for the  impurity resistivity (see \cite{Hewson}
and references theirin):
\begin{equation}
  R_{imp}(T) = \frac{R_0}{2} \left[ 1 - \frac{ \ln (T/T_K) } 
       { \left[ \ln^2 (T/T_K) + \pi^2S(S+1) \right]^{1/2} } 
 \right]
\end{equation} 
where $T_K$ corresponds to the Kondo scale, $T_K \sim W e^{-W/J}$
and $W$ is the bandwidth.
Assuming a power-law, $T^n$, for the resistivity of the conduction electrons,
the resistivity minimum will scale as 
\begin{equation}
    T_C \sim  J^{1/n}
\end{equation}
at weak couplings. Given the above, our QMC result $T_C \sim J$ follows by 
setting $n=1$ and we are left with searching for mechanisms which will 
lead to a linear in $T$ resistivity. A possibility is scattering on 
spin fluctuations \cite{Monthoux94,Moriya90,Hlubina95}. Note that both the KLM 
\cite{Capponi00} and DKLM an enhancement of 
local antiferromagnetic 
spin-spin correlations between impurity spin and conduction electron
($ \langle \vec{S}^{f}_{\vec{R}} \cdot \vec{S}^c_{\vec{R}} \rangle $)
is observed at an energy scale tracking $T_C$. Hence, magnetic fluctuations
are potentially present at this energy scale. 
Clearly further work is required to justify this hand waving argument.
\footnote{Note that in comparison to 
the KLM \cite{Capponi00}, $T_C$  is enhanced in 
the DKLM. Depleting further the DKLM of impurity spins, we should 
ultimately arrive to the single impurity case were $T_C$ scales to infinity 
when the conduction band is modeled by free electrons.}

Below $T_C$ the resistivity grows as a function of decreasing temperature. 
For values of $J/t > 0.8$, the scale  $T^{\star}$ emerges. 
In particular and for the considered couplings,
below  $T^{\star}$ the resistivity drops sharply, the specific shows a 
linear in $T$ behavior, and both the spin and charge susceptibilities seem to
converge to finite values. 
In a first approximation, this is what expects for
a Fermi liquid.  Given the limited data, it is hard to pin down the 
functional form of $T^{\star}$ as a function of $J/t$. Nevertheless, an 
exponential fit ($ e^{ - 1/\alpha J} $) accounts  reasonably well for the data. 
Due to the particle hole  symmetry of the model which automatically leads
to nesting, we cannot follow the metallic state to $T=0$. At $T_S$ substantial 
spin-spin fluctuations set in and drive the system to an insulator with activated 
resistivity. 
For values of $J/t < 0.8$, $T^{\star}$ drops below $T_S$.  Hence 
after a slow increase of resistivity below $T_C$,  an activated behavior   is 
expected below $T_S$.

\subsection{Comparing $T^{\star}$  to the single impurity Kondo 
           temperature}
\label{TKI} 
It is now interesting to compare $T^{\star}$ with the Kondo 
temperature of the single impurity problem:
\begin{equation}
      H_I = -t \sum_{\langle \vec{i},\vec{j}  \rangle, \sigma } 
 \left(  c^{\dagger}_{\vec{i}, \sigma}   
         c_{\vec{j}, \sigma} + {\rm H.c.} \right) + 
     J \vec{S}^{c}_{\vec{I}} \cdot 
    \vec{S}^{f}_{\vec{I}}.
\end{equation}
To obtain the Kondo temperature one can compute the impurity spin susceptibility:
\begin{equation}
  \chi_I = \int_{0}^{\beta} {\rm d } \tau \langle \vec{S}^{f}_{\vec{I}}(\tau) 
\vec{S}^{f}_{\vec{I}}(0)  \rangle
\end{equation}
where $ \vec{S}^{f}_{\vec{I}}(\tau) = e^{\tau H_I}  \vec{S}^{f}_{\vec{I}} 
e^{-\tau H_I} $. In the low temperature limit, $T_K^I$ is the only scale hence 
$T \chi_I = F(T/T_K^I)$ where $F$ is a universal function.  
\begin{figure}
\includegraphics[width=0.4\textwidth]{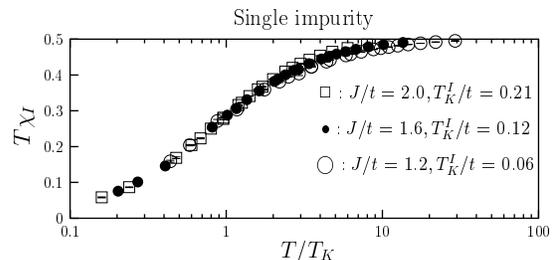}
\caption[]{Impurity spin susceptibility for the single impurity problem. 
The calculations were done with the Hirsch-Fye impurity  algorithm 
\cite{HirschFye86}. For the 
conduction electrons we considered a $512\times 512$ lattice.
}
\label{TK_imp.fig}
\end{figure}
In Fig. \ref{TK_imp.fig}
and for  the three considered values of $J/t$ a good collapse  of the data is 
obtained with the listed values of the Kondo temperature $T_K^{I}$. Comparison
with Fig. \ref{Scales_MF.fig} shows that the mean-field approximation produces 
next to identical results for the Kondo temperature. Recall that the Kondo
temperature for the impurity and lattice cases are identical in the mean-field 
approach. 

The important point however, is that within our accuracy 
$T^{\star}$ compares very well with $T_K^I$. This stands in rough agreement 
with experiments on Ce$_x$ La$_{1-x}$ Cu$_6$ \cite{Sumiyama86}.  
Here the single impurity
Kondo temperature ($x << 1$) is given by $T_K^I \sim 3 K$. As $x$ grows
($0.73 < x < 1 $), a 
maximum  in the resistivity versus temperature curve develops at energy scales 
in the range  $ T^{\star} \sim 5-15 K$.  It is also worth pointing out that 
below  $T^{\star}$ the spin susceptibility in  CeCu$_6$  seems to saturate
\cite{Sumiyama86}. This
stands in rough agreement with our results. 
Hence, $T^{\star}$ corresponds essentially to the single impurity Kondo 
temperature $T_K^I$  below which the magnetic impurities are screened. It marks 
the onset on the heavy-electron state. 

\section{Conclusions}
To summarize, we have presented a detailed numerical study of a 
depleted Kondo lattice. The depletion considered here is very special, 
and corresponds, to the best of our knowledge, to the minimal regular 
depletion  required to obtain a two-dimensional metallic state in the 
$J/t \rightarrow \infty $ limit.

From the technical point of view,  the QMC approach  is flexible and
can be applied to any depletion pattern, random or regular. The 
constraint to avoid the minus sign problem  
is the particle-hole symmetry of the conduction band. 
Hence various subjects may be studied. 
As function of depletion concentration, a metal-insulator transition 
should occur as observed in Ce$_3$Bi$_4$Pt$_3$ when Ce is replaced 
by La \cite{Hundley90}. In the strong coupling limit one expects 
this metal-insulator
transition to occur at the percolation threshold. Furthermore, in the
insulating phase random depletion of the $f$-sites  should trigger 
magnetic ordering between the magnetic moments induced by the depletion. 
Hence order by disorder is equally a phenomena which can studied in this
model.  To control size 
effects, we have introduced a magnetic field  perpendicular to the 
two dimensional lattice. We have shown that by scaling the magnetic 
field as $B L^2/\Phi_0$ extremely good size scaling to the thermodynamic 
limit is obtained. In 
particular already on small lattices the temperature scale 
at which size effects set in 
drops by an order of magnitude when the magnetic field is included. This 
is the key to our results since it opens the energy window at which 
coherence effects may be studied. 

The QMC results for the DKLM  show three temperature scales 
which are summarized in Fig. 
\ref{Scales_qmc.fig}. i) The largest scale, $T_C$,  
corresponds to the minimum in the resistivity 
and tracks $J/t$. ii) $T_S$ corresponds to the energy scale below
which the magnetic length scales grows presumably  exponentially with 
decreasing temperature. Due to
particle-hole symmetry and associated nesting properties those spin 
fluctuations drive the system to an insulating state. The ground state 
of the model remains a magnetically ordered insulator irrespective  of the 
value of $J/t$. As in the KLM \cite{Capponi00} our data supports 
$T_S \sim J^2$ at weak couplings, in accordance with the RKKY interaction. 
In the strong 
coupling limit $T_S$ is exponentially suppressed and vanishes at $J/t = \infty$.
Note that in the absence of nesting 
one would expect the occurrence of  a quantum phase transition between 
magnetically ordered and paramagnetic metallic states at a finite 
values of $J/t$.  This  is 
unfortunately not accessible to the QMC method due to severe sign problems.
iii) The temperature scale $T^{\star}$ as defined by the maximum in the 
resistivity versus temperature cure becomes apparent when it 
exceeds the  spin scale $T_{S}$. This happens at $J/t > 0.8$.   
In the temperature range $ T^{\star} < T < T_S$  we see 
signatures of Fermi liquid character: the specific heat is roughly linear 
in $T$ and the spin and charge susceptibilities appear to converge to finite values.
Unfortunately, we cannot follow this metallic phase  down to $T=0$ because 
of the onset of spin fluctuations below $T_S$. 
iv) We have computed the corresponding single impurity Kondo temperature ($T_K^I$)
with the Hirsch-Fye QMC algorithm. Within the considered small intermediate 
coupling range, our results suggest $T_K^I \sim T^{\star}$.

Comparison with the mean-field results (see Fig. \ref{Scales_MF.fig})
shows that the spin scale is well reproduced. Clearly, the
$T_C$ scale is absent at the mean-field level. The mean-field 
Kondo temperature $T_K$ is independent on the impurity density and hence is 
identical to that of the single impurity model. Comparison with {\it exact} 
QMC results shows that this mean-field approximation yields  next to exact
values of the single impurity Kondo temperature. Hence the mean-field Kondo
temperature which marks a phase transition in  this approximation evolves 
to the crossover scale $T^{\star}$ when the model is solved exactly.  

Finally we comment on the coherence scale. At first glance  it is tempting
to associate the coherence scale with $T^{\star}$ since signs of 
Fermi liquid character start emerging at this energy scale. 
However  this identification will lead 
to contradictions. In particular, in the strong coupling limit  the 
coherence scale scales to a finite value  
\footnote { In the strong coupling limit, the model reduces to the three band 
free electron model with Fermi line plotted in Fig. \ref{latt.fig}c.  
The coherence temperature 
is then given by the coherence temperature of the free electron model which 
is finite. }
but $T^{\star}$ is expected to track $T_K^I \propto J$.
Hence we can only speculate that the coherence scale 
-- as in the mean-field approach -- lies well below  
$T^{\star}$ and is masked by $T_S$.

We thank HLRS (Stuttgart) for generous allocation of CPU time on the Cray-T3E. 
I  acknowledge useful discussions  with 
S. Capponi, M. Feldbacher and A. Muramatsu. The Deutsche Forschungsgemeinschaft 
(DFG) is thanked for financial support under the grant number AS 120/1-1.


\end{document}